\documentclass[sigconf,nonacm]{acmart}
\renewcommand\footnotetextcopyrightpermission[1]{} 
\AtBeginDocument{%
  \providecommand\BibTeX{{%
    \normalfont B\kern-0.5em{\scshape i\kern-0.25em b}\kern-0.8em\TeX}}}

\usepackage{multirow}
\usepackage{graphicx}
\usepackage{bbm} 
\usepackage{bm} 
\usepackage{subcaption}
\usepackage{enumitem}

\newcommand{\ba}{\mathbf{a}}

\newcommand{\bx}{\mathbf{x}}

\newcommand{\bz}{\mathbf{z}}
\newcommand{\btcvae}{Mult-TCVAE }
\newcommand{\bvae}{Mult-$\beta$-VAE }
\newcommand{\calI}{\mathcal{I}}

\newcommand{\bmu}{\bm{\mu}}
\newcommand{\bsigma}{\bm{\sigma}}
\newcommand{\expect}{\mathbb{E}}
\newcommand{\Normal}[1]{\mathcal{N}(#1)}

\newcommand{\topk}{\operatorname{top_n}}
\newcommand{\countfn}{\operatorname{Count}}
\newcommand{\rmetric}{\operatorname{R-MET}}
\newcommand{\genre}{\text{Genre}}
\newcommand{\dctrl}{\delta_{\text{ctrl}}}
\newcommand{\dirrel}{\delta_{\text{irrel}}}
\newcommand{\basecorr}{\text{Corr}}
\newcommand{\ctrlcorr}{\text{RandCorr}}
\newcommand{\ctrlcorrbl}{\text{Corr}_{\text{ctrl}}}
\newcommand{\ctrlcorrct}{\text{Corr}_{\text{rand}}}
\newcommand{\easycorrbl}{\text{EasyCorr}_{\text{ctrl}}}
\newcommand{\easycorrct}{\text{EasyCorr}_{\text{rand}}}
\newcommand{\diffcorrbl}{\text{DiffCorr}_{\text{ctrl}}}
\newcommand{\diffcorrct}{\text{DiffCorr}_{\text{rand}}}

\begin{document}

\title{Controllable Recommenders using Deep Generative Models and Disentanglement}

\author{Samarth Bhargav}
\email{s.bhargav@uva.nl}
\orcid{0000-0001-5204-8514}
\affiliation{%
  \institution{IRLab, University of Amsterdam}
  \country{The Netherlands}
}

\author{Evangelos Kanoulas}
\email{e.kanoulas@uva.nl}
\orcid{0000-0002-8312-0694}
\affiliation{%
  \institution{IRLab, University of Amsterdam}
  \country{The Netherlands}
}

\renewcommand{\shortauthors}{Bhargav et al.}

\begin{abstract}
In this paper, we consider \textit{controllability} as a means to satisfy dynamic preferences of users, enabling them to control recommendations such that their current preference is met. While deep models have shown improved performance for collaborative filtering, they are generally not amenable to fine grained control by a user, leading to the development of methods like deep language critiquing. We propose an alternate view, where instead of keyphrase based critiques, a user is provided `knobs' in a \textit{disentangled} latent space, with each knob corresponding to an item aspect. Disentanglement here refers to a latent space where generative factors (here, a preference towards an item category like genre) are captured \textit{independently} in their respective dimensions, thereby enabling \textit{predictable} manipulations, otherwise not possible in an entangled space. We propose using a (semi-)supervised disentanglement objective for this purpose, as well as multiple metrics to evaluate the controllability and the degree of personalization of controlled recommendations. We show that by updating the disentangled latent space based on user feedback, and by exploiting the generative nature of the recommender, controlled and \textit{personalized} recommendations can be produced. Through experiments on two widely used collaborative filtering datasets, we demonstrate that a controllable recommender can be trained with a slight reduction in recommender performance, provided enough supervision is provided. The recommendations produced by these models appear to both conform to a user's current preference and remain personalized. 
\end{abstract}



\maketitle

\section{Introduction}

Auto-encoder based architectures have shown impressive performance in collaborative filtering with implicit feedback \cite{liang2018variational, ma2019learning, shenbin2020recvae}. However, these methods may fail to model short term or dynamic preferences of a user. Subsequently, methods to tackle this explicitly problem have been proposed, for example, through conversations via a conversational recommender system \cite{jannach2020survey}, or critiquing recommendations using language or keyphrases \cite{gao2019dlc, li2020rankopt, luo2020deepcritiquing, yang2021bayesiancrit}. In contrast to keyphrases, typically mined from reviews or descriptions (for instance), this work considers building controllable or critique-able recommenders using attribute data for items, which can be used to construct preference distributions at a user level. This is in turn used in the critiquing process or to express a short-term preference, allowing a user to exert control over a recommender in a meaningful manner. For instance, a user who typically watches \textit{Action} movies, or (as a consequence) exclusively receives \textit{Action} recommendations, can now explicitly request \textit{personalized} movies of other genres, like \textit{Animation}. 
 
Controllability can be achieved by utilizing the \textit{generative} nature of certain \emph{Deep Generative} Recommenders \cite{gao2019dlc, li2020rankopt, luo2020deepcritiquing, yang2021bayesiancrit}. Such models have an encoder which produces a user-latent representation, which is fed to the decoder to predict items for that user. A manipulation of this representation, followed by decoding step using this, should alter recommendations produced by the model. However, since the latent spaces of these models are typically \emph{entangled}, the outputs produced through such manipulations are likely to be unpredictable or random, making them unusable for this task. 

We propose using a \textit{disentangled} latent space which, in contrast, can be manipulated predictably. A representation is disentangled w.r.t known ground truth variables or generative factors (ex. genres, availability, context, etc), if there is \textit{only one} latent dimension in the representation that is influenced when this ground truth variable is changed \cite{bengio2013representation, higgins2017beta, locatello19a, locatello2019disentangling}.

Prior work tackles critiquing/controllability by modeling a latent space where keyphrases are \textit{co-embedded} with user/item embeddings. By `zero-ing' out a certain keyphrase, the corresponding keyphrase embedding is used to update the user representation, producing critiqued recommendations. Our work in contrast doesn't utilize a co-embedding space, and instead the latent space is directly manipulated. This means no addition optimization (like in \cite{luo202latentlinear, li2020rankopt}) is required to incorporate (multi-step) critiques. Furthermore, this also allows for `positive' critiques as well as `soft' control (gradual, non-binary) instead of only binary critiques. 

We propose using supervision to obtain a mapping of a particular factor/aspect to a latent dimension. Since supervision signals might not be available for all users, we experiment with settings where limited data is provided.

In addition, while there have been metrics previously proposed for evaluating controllability/critiquing \cite{gao2019dlc, yang2021bayesiancrit, wang2021controllable}, these metrics don't explicitly account for the personalization of post-critique recommendations. We propose multiple metrics to evaluate the personalization of critiqued recommendations, measuring both binary critiques as well as `soft', gradual controllability. To summarize, our contributions are as follows \footnote{ \url{https://github.com/samarthbhargav/disentangled-control-recsys}}:

\begin{itemize}
    \item We propose using \textit{(semi-)supervised disentanglement} to learn disentangled representations for controllable, \textit{personalized} recommendations. Through experiments on two large scale collaborative filtering datasets using 2 types of signals as `knobs', we show the proposed models produce controllable recommendations, at the cost of a slight reduction in overall recommendation performance. In addition, we experiment with different levels of supervision, and show controllability can be achieved with limited data. 
    \item We focus on retaining recommendation performance while achieving effective controllability, and propose several metrics to extensively study (a) the degree of personalization of controlled recommendations (b) whether control is achieved in isolation (i.e if only one factor changes at a time) and (c) the effect of disentanglement on controllability. Using these metrics, we demonstrate that such models are amenable to user control, and the controlled recommendations appear to be personalized to an extent.
\end{itemize}

\begin{figure}[tp!]
    \centering
    \includegraphics[width=0.45\textwidth]{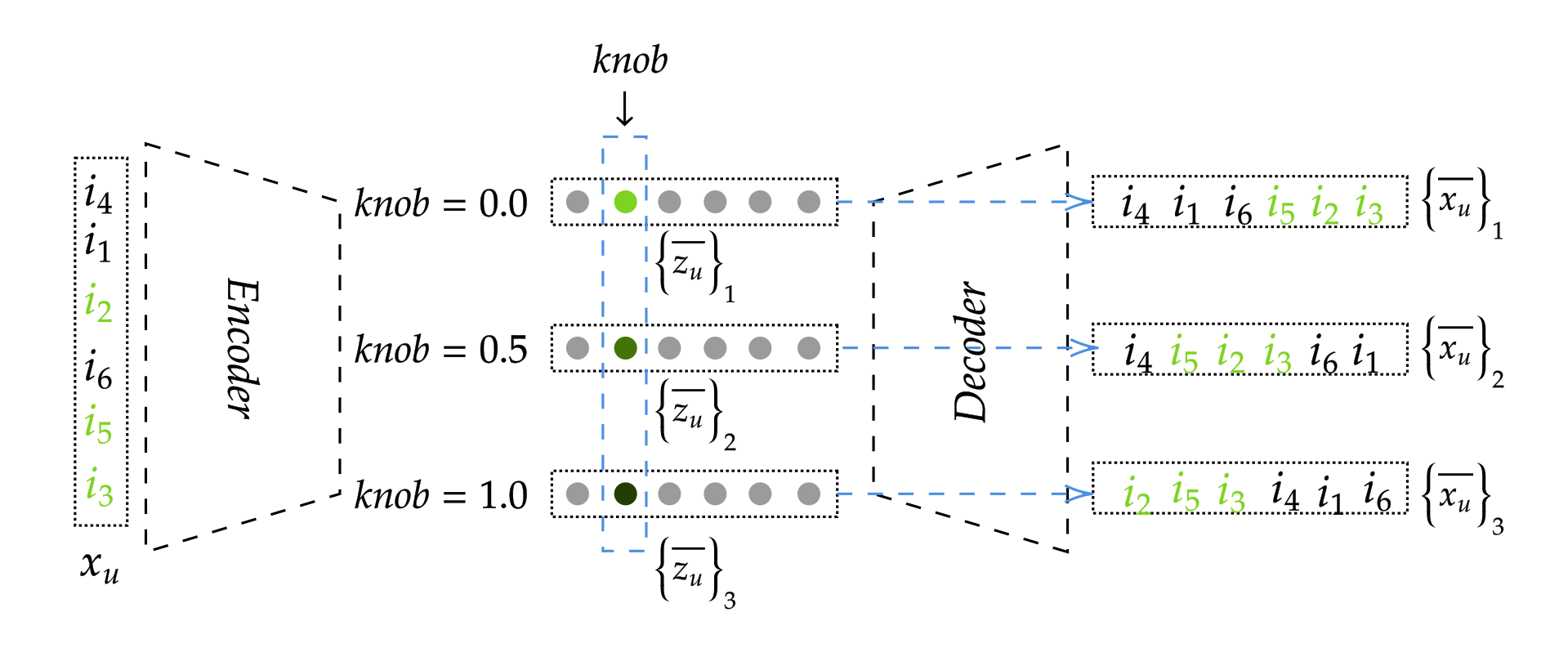}
    \caption{Deep Controllable Recommendations with (semi-)supervised disentanglement: A user's interactions are captured in $\bx_u$, used to infer the user representation $\bz_u$. Given a latent dimension $j$ (the `knob'), corresponding to a known factor (green items), only the values corresponding to the $j$th dimension of $\bz_u$ are replaced to produce $\{\bar{\bz}_u\}_{1,2,3}$. Each of these are then decoded to produce recommendations. The manipulation here pushes the green items higher in the ranked list, leading to controllable recommendations.}
    \label{fig:architecture}
\end{figure}

\section{Background and Related Work}

\subsection{Deep Recommender Systems}\label{sec:bg_deep_rec}

There have been several works that use Deep Learning for Recommendation \cite{wang2015collaborative,he2017neural,wang2019neural,zhang2019deep}. A common theme in several deep models is the use of auto-encoder architectures like the Collaborative Denoising Autoencoder (CDAE) \cite{wu2016collaborative} or MultVAE \cite{liang2018variational}. The latter uses the Variational Autoencoder framework \cite{blei2017variational, kingma2013auto} for recommendation. \citet{shenbin2020recvae} propose the state-of-the-art RecVAE. Deep Recommender systems however are (typically) black-box models which are difficult to interpret, compared to content-based methods \cite{zhang2019deep}.  

\subsection{Disentangled Representation Learning}\label{sec:bg_disent}

The use of deep \emph{generative} recommenders allow for disentangled representations, regaining some interpretability \cite{ma2019disentangled, wang2020disentangled, cui2020disentangled}. Most such models use the VAE framework, where the objective is to \emph{reconstruct} the input with high probability, with an additional loss term \textit{constraining the latent space}. By imposing additional constraints, \textit{disentangled} representations can be learnt. These methods are typically unsupervised and therefore come with limitations \cite{locatello19a}: they were found to be very sensitive to hyperparameters, \textit{reliably} learning unsupervised disentangled representations is a very challenging task. 

There have been several works that use disentanglement in recommendation: \citet{ma2019learning} assume that disentanglement is generated by user behaviours on `macro' and `micro' levels, and show that their models outperform non-disentangled baselines.  \citet{ma2020disentangled} apply disentanglement to the sequential recommendation task, while \citet{wang2020disentangled} disentangle diverse user-intents using graph based collaborative filtering; \citet{cui2020disentangled} propose DGCF, which have `implicit' (unknown) and `explicit' (known) signals that are disentangled using an RNN based model with a two-step method since some computations are non-differentiable. \citet{wang2021controllable} propose using weakly-supervised disentanglement objective on pairs of items. This allows for attribute-based item retrieval, such that items differ more/less on the provided attribute. In contrast to prior work, we use semi-supervised disentanglement, while focusing on using disentanglement for controllability.

\subsection{Critiquing / Controllable Recommenders}\label{sec:bg_ctrl}

We focus on critiquing for \textit{deep} recommenders (for a full overview of earlier work in critiquing, see Chapter 13 of \cite{riccihandbook}). The following paragraphs detail approaches that leverage language or keyphrases for critiquing, followed by other approaches like ours which does not utilize any language/keyphrase. We end this sub-section with a discussion about evaluation metrics.

Recent work has focused on using language or keyphrase based critiquing \cite{gao2019dlc, li2020rankopt, luo2020deepcritiquing, yang2021bayesiancrit}. \citet{gao2019dlc} propose the Deep Language based Critiquing (DLC) paradigm, where (subjective) keyphrases are treated both as explanations as well as a means for a user to critique recommendations. Here, critiquing is achieved by rejecting or `zero'-ing out keyphrases which are co-embedded with users. \citet{luo202latentlinear} adapt the CE-NCF for \textit{multi-step} critiquing; \cite{li2020rankopt} use a ranking  optimization instead of pairwise re-scoring to re-rank items; \cite{luo2020deepcritiquing} use a VAE-based model to achieve better training stability and lower computational complexity; \cite{yang2021bayesiancrit} propose using Keyphrase Activation Vectors instead of using a second 'head' used in \cite{gao2019dlc}, and adapt it for `positive' critiques. \citet{antognini2020trecs} propose T-RECS, which infers keyphrases from the intersection of user profiles and an item. \cite{antognini2021fastmulti} show that a model under a weak-supervision scheme matches or beats recommendation/explanation and critiquing performance while being much faster. We use supervised disentanglement in the user-latent space instead of a separate embedding space, while not requiring any language data (at the cost of no explanations). Prior work focuses on binary signals ex. reject/accept, while our work considers `soft' critiques, allowing for a \textit{gradual} tuning. Note that while we experiment only with unit critiques, compound and/or multi-step critiques is also possible within our proposed models, which we leave for future work. 

\citet{wang2021controllable} propose an orthogonal task, where items are retrieved on a `gradient' given an attribute, focusing on \textit{gradual} change, similar to our work. Our work focuses on recommendation performance whereas \cite{wang2021controllable} focus on gradient item retrieval. \citet{cen2020controllable} propose ComiRec for sequential recommenders, where control is used to balance diversity and recommendation accuracy, from the perspective of a designer. The method presented in \cite{ma2019learning} is closest to us, where representations are first altered, followed by a nearest neighbours search over items using a beam search. In contrast, we utilize the \emph{generative nature} of the model instead of performing nearest neighbour search, in addition to using semi-supervised disentanglement. The authors are aware of concurrent work \cite{nema2021disentpref} that is close to ours. Since the paper/code was unavailable at the time of experiments/ publication, we leave the evaluation of \cite{nema2021disentpref} for future work. 

\subsubsection*{Metrics for critiquing/controllability} \citet{gao2019dlc} propose the Falling MAP for \textit{negative} critiques. Given a user and a keyphrase (here, genre/tag) this metric measures the difference in MAP values computed on the set of items that have the critiqued keyphrase. If this value is positive, items of the critiqued keyphrase `falls' down the ranking list, which is desirable. Similarly \cite{yang2021bayesiancrit} propose a follow up metric, which accounts for positive critiques as well, comparing the normalized difference of average ratings of items before / after the critique. We note that these metrics don't explicitly model the \textit{personalization} of post-critique recommendations. In contrast, the proposed metrics explicitly consider this, by measuring performance on against items for which a user's preference is known. 

\section{Methodology}\label{sec:meth}

We present disentanglement of the latent space using semi-supervision as a means to control in Section \ref{sec:meth_ctrl}. Adapting disentanglement representation learning for this task is explored in detail in Section \ref{sec:meth_disent}. 

\subsection{Control using (semi-)supervised disentanglement}\label{sec:meth_ctrl}

We train a model using a (semi-)supervised disentanglement loss, producing disentangled user representations. During inference, representations for all users are inferred (possibly offline). In the foreground, the representations are manipulated on the fly by increasing/decreasing a `knob', allowing for granular change. In the background, the mapping of a knob to a latent dimension is utilized to modify the user's representation, which is fed to the decoder to produce controlled recommendations. The following paragraphs motivate the use of disentanglement and semi-supervision to build such a recommender.

As mentioned before, the entangled nature of a typical deep recommender renders meaningful manipulation impossible. This is because (a) there is no guarantee that manipulations produce changes in a single generative factor, and (b) it is not apparent \textit{which} dimension to alter given a factor. Even if we assume (b), predictability is not guaranteed, ex. manipulating the \textit{Thriller} dimension may increase the number of \textit{Thriller} movies, but it may inadvertently increase the number of \textit{Children} movies, which is undesirable. 

We propose using supervised disentanglement to tackle these issues. Disentanglement can isolate a generative factor to a single dimension, and altering this dimension (assuming perfect disentanglement) space should produce a change only in this generative factor. Using \textit{(semi)-supervised} disentanglement addresses two issues: (a) unsupervised disentanglement has several shortcomings \cite{locatello19a}, which semi-supervision may alleviate; and (b) a particular dimension is now `tied' to a generative factor, ensuring that the correct dimension is being altered.

These recommendations should ideally be \textit{personalized}, since they were trained with a reconstruction loss. In essence, a recommender system can be tuned \textit{on the fly} to express user preferences for the current session. This process is illustrated in Figure~\ref{fig:architecture}. 

In this paper we opt for preference distributions i.e tag/genre distributions, as generative factors or `knobs' (see Section \ref{sec:exp_data_gen_factors}), since they are widely available. However, preference distributions are \textit{not independent} and can also be very \textit{noisy} \cite{msd-Bertin-Mahieux2011, movielens}, in contrast to independent and exact generative factors used in disentanglement literature e.g dSprites\cite{dsprites17}. Consequently, they may be difficult to learn and/or disentangle. The following section discusses particulars about the models.    
\subsection{Unsupervised and Semi-supervised Disentanglement}\label{sec:meth_disent}

In this section, two models, the $\beta$-VAE \cite{liang2018variational, higgins2017beta} and the $\beta$-TCVAE \cite{chen2018isolating}, are adapted for the recommendation task using a Multinomial loss \cite{liang2018variational}, resulting in recommenders that can produce disentangled representations. Furthermore, this objective is further supplemented with a semi-supervision loss.  

\subsubsection{Unsupervised Disentanglement} The VAE \cite{kingma2013auto} optimizes the marginal (log-)likelihood of the observed data $\bx$ in expectation over the whole distribution of latent factors $\bz$. It assumes that the data is supported on a low dimensional manifold in a high dimensional space. The assumption of a latent code allows us to express the marginal distribution as follows: $p_\theta(\bx) = \int_\bz p_\theta(\bx|\bz)p(\bz)$. Since this quantity is intractable, the Evidence Lower Bound is optimized instead \cite{jordan1999introduction}: 

\begin{equation}\label{eq:vae_loss}
    \log p_\theta(\bx) \geq \mathcal{L}_{VAE} = \expect_\bx [\expect_{q_\phi(\bz|\bx)} [\log p_\theta(\bx|\bz)] - KL[q_\phi(\bz|\bx) || p(\bz)]]
\end{equation}

\noindent where $q_\phi(\bz|\bx)$ is a variational approximation of the posterior distribution, which is typically a Gaussian distribution. $p(\bz)$, the prior, is usually an isotropic Gaussian: $p(\bz) = \Normal{\mathbf{0}, \mathbf{1}}$. Several methods modify this objective to encourage disentanglement:

\subsubsection{$\beta$-VAE}\label{sec:ex-beta-vae} The $\beta$-VAE\cite{higgins2017beta, burgess2018understanding} modifies the VAE loss i.e Equation \ref{eq:vae_loss} by adding a $\beta$-multiplier to the KL term:

\begin{equation}\label{eq:bvae_loss}
    \mathcal{L}_{\beta VAE} = \expect_\bx[\expect_{q_\phi(\bz|\bx)} [\log p_\theta(\bx|\bz)] - \beta . KL[q_\phi(\bz|\bx) || p(\bz)]]
\end{equation}

Disentanglement is encouraged by setting $\beta > 1$, which further forces the posterior distribution $q_\phi(\bz|\bx)$ to be close to the \textit{isotropic} prior for every $\bx$ (as opposed to only on average), imposing statistical independence. However, this involves a trade-off between reconstruction quality and disentanglement \cite{higgins2017beta, burgess2018understanding}. While \citet{liang2018variational} use $\beta<1$ for better recommender performance, we opt for $\beta>=1$ to encourage disentanglement, and achieve similar/better recommender performance by removing the Dropout layer, following \citet{shenbin2020recvae}. An annealing strategy is used for $\beta$, following \cite{burgess2018understanding, liang2018variational}. The resulting model, which uses a Multinomial loss (see Section \ref{sec:mult-loss}), is termed \bvae.

\subsubsection{$\beta$-TCVAE}\label{sec:ex-btc-vae} \citet{chen2018isolating} show that by imposing a constraint on only a part of the KL divergence term, reconstruction and disentanglement increases compared to the $\beta$-VAE model. The KL term is decomposed into 3 terms (a) an index-code Mutual Information (MI) term, (b) dimension-wise KL, and (c) a total correlation (TC) term. The index-code MI captures the information between the data and latent variables, based on the empirical data distribution; the dimension-wise KL prevents deviation of the posterior from the prior dimensions; and the TC term encourages statistical independence of the posterior distribution leading to disentanglement. The key idea is that the $\beta$-VAE constrains all 3 terms, which might harm performance, while only the TC term is penalized in this model. The loss function of the TC-VAE is as follows: 

\begin{align}
    L_{\beta-TCVAE} &= \expect_\bx[\expect_{q_\phi(\bz|\bx)}[\log p_\theta(\bx|\bz)] - \alpha I_q(\bz;n) \nonumber\\
                    &- \gamma \sum_j KL[q(\bz_j) || p(\bz_j)] - \beta KL[q_\phi(\bz) || \Pi_j q_\phi(\bz_j)] ]\label{eq:btcvae-loss}
\end{align}

where $\alpha$, $\gamma$ and $\beta$ are multipliers for the index code MI, the dimension wise KL and the TC term respectively. Following \citet{chen2018isolating}, $\gamma = \alpha = 1$ is used, $\beta$ is a hyperparameter.  The final model, which we term \btcvae, uses a Multinomial loss (see Section \ref{sec:mult-loss}).

\subsubsection{Semi-supervised Disentanglement} The distribution $q_\phi(\bz|\bx)$ can be constrained so that a given dimension of $q(\bz)$ is associated with a ground truth attribute. That is, given a representation (for instance, the mean vector from the posterior) for a data point, $\bmu \in \mathbb{R}^D$, and the set of known attributes $\ba \in \mathbb{R}^A$ for that data point,  with $A \leq D$, the $j$th dimension of $\bmu$ is predictive of the $j$th dimension of $\ba$, $1 \leq j \leq A$. This is achieved by adding a semi-supervised loss $R_s$ :

\begin{equation}\label{eq:ss_loss}
    \mathcal{L}_{ss} = \mathcal{L}_{\text{unsup}} + \gamma_{ss} .  \expect_{\bx,\bz} R_s(q_\phi(\bz | \bx), \ba)
\end{equation}

$\mathcal{L}_{\text{unsup}}$ is either Equation \ref{eq:bvae_loss} or \ref{eq:btcvae-loss}. The binary cross entropy loss is used for $R_s$ \cite{locatello2019disentangling}:

\begin{equation}\label{eq:bce_loss}
    R_s(\bmu, \ba) = \sum_{i=1}^A \ba_i \log(\sigma(\bmu_i)) + (1 - \ba_i) \log(1 - \sigma(\bmu_i)) 
\end{equation}

where $\sigma(.)$ is the logistic function. Note that the remaining dimensions of $\bz$, those that are not supervised, are not constrained by the semi-supervision loss.

\section{Evaluation Methodology}\label{sec:eval}

In this section we propose an evaluation methodology and a set of metrics that investigates the extent of controllability and personalization of a recommender, described in Section \ref{sec:eval_ctrl}. We also briefly describe existing metrics for evaluating disentanglement in Section \ref{sec:eval_disent}. 

\subsection{Evaluating Controllability}\label{sec:eval_ctrl}

\begin{table}[tpb]
\centering
\caption{Metrics for Controllability, listed with the inputs and holdouts used in its computation. $g_j$ is the genre being manipulated, while $g_i$ is another genre}
\label{tab:ctrl_metrics}
\resizebox{0.3\textwidth}{!}{%
\begin{tabular}{@{}lll@{}}
\toprule
Metric & Input & Holdout(s) \\ \midrule
$\dctrl$ & $\calI_u - \calI_{(u, g_{j})}$ & $\calI_{(u, g_{j})}$ \\
$\dirrel$ & $\calI_u - \calI_{(u, g_{j})}$ & $\calI_{g_j} - \calI_{(u, g_j)}$ \\
$\basecorr$ & $\calI_u - \calI_{(u, g_{j})}$ & $\calI_{(u, g_{j})}$ \\
$\ctrlcorrbl$ & $\calI_u - \calI_{(u, g_{j})} - \calI_{(u, g_i)}$ & $\calI_{(u, g_{j})}$ \\
$\ctrlcorrct$ & $\calI_u - \calI_{(u, g_{j})} - \calI_{(u, g_i)}$ & $\calI_{(u, g_{i})}$ \\ \bottomrule
\end{tabular}%
}
\end{table}

For ease of discussion, this section assumes a single factor $g_j$, s.t given a user $u$, and a factor $g_j$, the set of all items with factor $g_j$ is $\calI_{g_j}$, items encountered/rated by a user is $\calI_u$,  and items with factor $g_j$ that user $u$ has encountered is $\calI_{(u, g_j)} = \calI_{g_j} \cap \calI_{u}$ (subscripts for $\bz$ are dropped for brevity in the following descriptions). Note that an item can have multiple factors e.g a movie can be both \textit{Action} and \textit{Adventure}.

Given a user representation $\bz$, the dimension corresponding to $g_j$ is obtained by using a mapping function $k$, s.t $k(g_j)$ is the dimension corresponding to $g_j$ in $\bz$, with $k(g_j) \in \{1, \dots K\}$, with $K \leq D$, where D is the dimensionality of the latent space. In addition, we assume that manipulations of each dimension are in $\in [0, 1]$ (0 is minimum). A manipulation entails the replacement of the $k(g_j)$ dimension with a value indicating the position of a `knob', producing $\bar{\bz}$. In addition, the top-n items of $\bar\bx$ is denoted by $\topk(\bar\bx)$, and the number of items of a factor $g_j$ in the top-n list is denoted by $\countfn(\topk(\bar\bx), g_j)$. Finally, the $i^{th}$ element of a vector $\ba$ is denoted by $[\mathbf{b}]_i$, with $[\mathbf{b}]_i = 0$ indicating that the $i^{th}$ dimension is set to 0.

\subsubsection{Desiderata}
For evaluating controllability, there are a list of desiderata: 

\begin{enumerate}
    \item Increasing the values of the $k(g_j)$ dimension should push items of $g_j$ higher. In other words, given $[\bar{\bz}^m]_{k(g_j)} > [\bar{\bz}^n]_{k(g_j)}$ and assuming $\bar{\bz}^m$ produces $\bar{\bx}^m$ and $\bar{\bz}^n$ produces $\bar{\bx}^n$ respectively, then: $$\countfn(\topk(\bar{\bx}^m), g_j) > \countfn(\topk(\bar{\bx}^n), g_j)$$ for small values of $n$. A natural consequence of this is irrelevant items $\calI{g_{-j}}$ should be pushed down or be replaced with relevant items, i.e.\ $\countfn(\topk(\bar{\bx}^m), g_{-j})$ should decrease. Put another way, $k(g_j)$ should only control items belonging to $g_j$, and \emph{not} control items of other factors.   
    \item In addition, it is crucial to ensure that these recommendations are \textit{personalized} i.e $\topk(\bar{\bx}^m)$ should have items of the requested genre, but only those items that the user might like, i.e.\ items in $\calI_{(u, g_j)}$. 
\end{enumerate}

We assume a ranking metric $\rmetric$, computed against a held out set of items. $\rmetric$ can be any ranking metric, e.g.\ NDCG, assuming the relevance of all items except the ones in the holdout set to zero. We consider five metrics in total, with the two $\delta$ metrics, $\dctrl$ and $\dirrel$, measuring the ranking changes of items of interest versus other items; and three correlation metrics  measuring the gradual change observed as a dimension is manipulated: $\basecorr$, and ($\ctrlcorrbl$ ,$\ctrlcorrct$). The $\delta$ metrics measures the change induced when a user prioritizes items of only one factor, reflecting a critique or a short term preference i.e `show me \textit{Animation} movies only'. The correlation metrics on the other hand can capture exploratory behaviours or soft preferences (`maybe I will try at \textit{Horror} movies'), and measures to what extent the recommender is \textit{interactive and predictable}. These metrics are summarized in Table \ref{tab:ctrl_metrics} along with their inputs and holdout sets. Note that all the metrics are averaged across genres and then users.

\subsubsection{$\delta$ metrics}

Similar to $F\mbox{-}MAP$ \cite{gao2019dlc} and \textit{PostCritRatingDiff} \cite{yang2021bayesiancrit}, \textbf{$\dctrl$} measures the change in the ranking as a knob is set to its maximum value. Note, however, that only items \textit{not} belonging to $g_j$ that the user likes are fed into the model, with the intent to observe if manipulating the $k(g_j)^{th}$ dimension increases the positions of the held-out items of genre $g_j$ \textit{that a user likes}. As a consequence, the model has to infer the correct items of $g_j$ to recommend, without knowledge of the users preference towards items of $g_j$. Furthermore, in contrast to prior metrics, this evaluation is \textit{personalized} i.e the metric is computed against $\calI_{(u, g_{j})}$ and not $\calI_{g_{j}}$. The process to compute the metric is detailed below.

First, \textit{before} the manipulation, items of \textit{other} genres that the user has liked, $\calI_u - \calI_{(u, g_{j})}$ is used to infer $\bx$. $\rmetric_{\text{default}}$ is computed on this, using $\calI_{(u, g_j)}$ as the holdout set, measuring the baseline ranking score of \textit{personalized} items belonging to $g_j$. Second, $[\bar{\bz}]_{k(g_j)}$ is set to max indicating a high preference, and decoded to produce $\bar{\bx}$. This is used to compute $\rmetric_{\text{ctrl}}$ on the same holdout, expressing the ranking score of \textit{personalized} items belonging to $g_j$ \textit{after} manipulation. The difference between the two metrics is $\dctrl$:  

\begin{equation}
    \dctrl = \rmetric_{\text{ctrl}}  - \rmetric_{\text{default}}
\end{equation}

The higher the $\dctrl$, the better the ability to produce personalized and controlled recommendations. 

The second metric, \textbf{$\dirrel$}, is a variation of $\dctrl$. While recommending items of genre $g_j$ before items of other genres is important, it is desirable to recommend \textit{items that the user might like}, over irrelevant items of $g_j$. $\dirrel$ is similar to $\dctrl$, except the holdout set are items of $g_j$ \textit{not} rated by the user, $\calI_{g_j} - \calI_{(u, g_j)}$. If $\dirrel$ is high, items that the user \emph{might not like} are also being promoted, which may be undesirable. While these two metrics reflect the extent of controllability, granular metrics, reflecting \textit{gradual} change is presented in the following section.

\subsubsection{Correlation Metrics} The following metrics compute the correlation between changes in ranking metric $\rmetric$ and the \textit{values} of the manipulated dimension. The correlation measure used in the experiments is the Pearson's correlation, we leave non-linear variants to future work.



\textbf{$\basecorr$} quantifies the correlation of \textit{gradual} manipulations (as opposed to setting it to the max value as with $\dctrl$) with $\rmetric$ using $\calI_{(u, g_j)}$ as the holdout set. \textbf{$\ctrlcorrbl$} and \textbf{$\ctrlcorrct$} measure the same correlation with different holdout sets and identical inputs. \textbf{$\ctrlcorrbl$} uses $\calI_{(u, g_{j})}$ as the hold out, measuring the effect of manipulation against items of $g_j$, whereas \textbf{$\ctrlcorrct$} uses $\calI_{(u, g_{i})}$ which measures the effect of manipulation against items of a random \textit{control} factor $g_i$. The input set for the last 2 metrics has neither items of $g_j$ \textit{nor} $g_i$. In the ideal case, $\ctrlcorrbl$ should be positive i.e controlling $k(g_j)$ positively influences items of $g_j$, but $\ctrlcorrct$ should be zero (no change) or negative (decrease, either due to replacement or demotion), meaning items of another factor are not/negatively influenced.  

It is our hypothesis that keeping items of irrelevant genres low in the ranking is harder if the genre being controlled is highly correlated with another genre in the data (e.g many \textit{Action} movies are also tagged \textit{Adventure}). To measure this, we use two $g_i$: $g_{\text{easy}}$ and $g_{\text{difficult}}$, where $g_{\text{easy}}$ is the least co-occurring factor with $g_j$ and $g_{\text{difficult}}$ conversely frequently co-occurs with $g_j$ producing $\easycorrbl$, $\easycorrct$, $\diffcorrbl$ and $\diffcorrct$ \footnote{Co-occurrences are computed at user level and may differ across users}.

\subsection{Evaluating Disentanglement}\label{sec:eval_disent}

In addition to controllability, disentanglement of the user representations can also be evaluated since the ground truth generative factors are available. Several metrics have been proposed to evaluate disentanglement, such as the Mutual Information Gap (MIG) \cite{chen2018isolating}, the $\beta$-VAE metric \cite{higgins2017beta} or the FactorVAE metric \cite{kim2018disentangling}. We evaluate disentanglement with the Mutual Information Gap (MIG) \cite{chen2018isolating}, due to ease of implementation, wide applicability and the unbiased nature of the metric for all hyperparameter settings \cite{locatello19a}. 
Given a generative factor, the empirical mutual information (M.I) between the values of the ground truth generative factors and each dimension of encoded samples from the data is computed. The MIG of this generative factor is then the difference between the highest and second highest MI values. This quantity is averaged across all generative factors to obtain the MIG score. A MIG of 1 for a particular generative factor implies that one latent dimension has MI=1, with others having MI=0.

\section{Experiments}\label{sec:exp}

The datasets, along with the requisite processing steps and generative factors for supervision are described in Section \ref{sec:datasets}. The research questions are described in Section \ref{sec:exp_rqs}, followed by specifics of models in Section \ref{sec:exp_setup}.

\subsection{Datasets}\label{sec:datasets}

In our experiments we use the Million Songs Dataset (MSD)\cite{msd-Bertin-Mahieux2011} and the Movielens-20M (ML-20M) \cite{movielens}, two widely used collaborative filtering datasets. The steps for preparing the dataset for training are outlined in Section \ref{sec:exp_data_proc}. Preference distributions are constructed as signals for supervision and for evaluating disentanglement/controllability, which is described in Section \ref{sec:exp_data_gen_factors}.

\subsubsection{Dataset processing}\label{sec:exp_data_proc} Each dataset is processed according to the steps outlined in \cite{liang2018variational}. The users are split into test, validation and train sets. The test/validation set size for ML-20M is 10,000 users and for MSD is 50,000 users, with 20\% of items held-out. The models is trained with the entire train history, and evaluated with the 20\% held out set. The data is binarized by keeping ratings of four or higher and only users who have interacted with at least 5 movies are kept.

\subsubsection{Generative Factors}\label{sec:exp_data_gen_factors} The generative factors considered for both datasets are \emph{preference distributions} computed for each user using genres (MSD/ML-20M) or tags (MSD only). The number of users, items, generative factors are reported in Table \ref{tab:datasets}. For a user $u$, $\genre_u$ is computed by calculating the proportion of movies that belong to a genre (or tag), divided by the total number of movies that the user has watched, capturing the (global) preferences of a user towards a genre/tag. Since there are $522362$ tags in MSD with many repeats, only the most frequent are picked and grouped together before computing the distribution\footnote{Due to a lack of space, these are not included in the paper, and can be accessed at \texttt{datasets/msd/selected\_tags.tsv}}. The dimensions of the signals for ML-20M is 19, and for MSD (Genre) is 21, and for MSD(Tag) is 30, with the supervision constraining only the first few dimensions.

\begin{table}[t]
\caption{Datasets used, along with signals being considered. The last column reports the MIG score, computed on the ground truth labels against itself using 10,000 users sampled from the test set}
\label{tab:datasets}
\resizebox{0.45\textwidth}{!}{%
\begin{tabular}{@{}lllll@{}}
\toprule
Dataset (ID) & \# users & \# items & Generative Factors               & G.T MIG \\ \midrule
ML-20M       &    136,677     &   20108       & genre distribution                       &        0.2019     \\
MSD (Genre)    & 571,355 &   41140       & genre distribution                         &           0.3238  \\
MSD  (Tag)  & 571,355 &   41140       & tag distribution                         &   0.2525          \\ \bottomrule
\end{tabular}%
}
\end{table}

\subsection{Research Questions}\label{sec:exp_rqs}

This section details the experimental setup employed in the paper. The evaluation metric for recommender performance throughout this paper is NDCG@100, following \cite{liang2018variational, shenbin2020recvae} i.e $\rmetric$ is NDCG for all experiments. Controllability for models with 0 supervision cannot be evaluated since $k(.)$ is unavailable. In addition, the manipulations (which replaces $[\bar{\bz}]_{k(g_j)}$) are produced by using the Inverse CDF of the probability distribution used during training \cite{kingma2013auto}.   

\newcommand{\RQone}{How much supervision is needed for achieving controllability? How does controllability vary across datasets and models?}
\newcommand{\RQtwo}{To what extent does disentanglement affect controllability of models? }
\newcommand{\RQthree}{What effect does introducing controllability into a model have on recommendation performance?}

\begin{table*}[htbp]
\caption{Results for ML-20M: MIG measures disentanglement, $\dctrl$ and $\basecorr$ measure control, with $\easycorrbl$, $\easycorrct$, $\diffcorrbl$ and $\diffcorrct$ contrasting control of one genre against another. NDCG and $\dirrel$ (when contrasted with $\dctrl$) measure personalization of non- and controlled recommendations respectively. Values in braces are standard errors.}
\label{tab:res_ml20m}
\resizebox{\textwidth}{!}{%
\centering
\begin{tabular}{ccc|cc|cc|c|cc|cc}
\toprule Model & \% & $\beta$ & NDCG &  MIG & $\uparrow$ $\dctrl$ & $\downarrow$ $\dirrel$ & $\uparrow$ $\basecorr$ & $\uparrow$ $\easycorrbl$ & $\downarrow$ $\easycorrct$ & $\uparrow$ $\diffcorrbl$ & $\downarrow$ $\diffcorrct$ \\ \midrule
\multirow{3}{*}{\rotatebox{30}{\bvae}} & 0 & 0 & 0.2782 & 0.0016 & - & - & - & - & - & - & -  \\
 & 50 & 0 & 0.2627  & 0.1265 & 0.1995 (0.0153) & 0.0658 (0.0100) & 0.8859 & 0.8880 (0.0045) & -0.2487 (0.0132) & 0.8547 (0.0055) & -0.5793 (0.0115) \\
 & 100 & 0 & 0.2587  & 0.1417 & \textbf{0.2246} (0.0190) & 0.0764 (0.0127) & 0.8628 & 0.8677 (0.0053) & -0.1994 (0.0119) & 0.8408 (0.006) & -0.5911 (0.0113) \\ \midrule
\multirow{3}{*}{\rotatebox{30}{\btcvae}} & 0 & 0 & \textbf{0.2904}  & 0.0007 & - & - & - & - & - & - & - \\
 & 50 & 0 & 0.2802 & 0.0688 & 0.2145 (0.0157) & 0.0705 (0.0104) & \textbf{0.8901} & 0.8907 (0.0044) & -0.2869 (0.0125) & 0.8599 (0.0053) & -0.6448 (0.0104) \\
 & 100 & 0 & 0.2791 & \textbf{0.1819} & 0.2210 (0.0198) & 0.0755 (0.0129) & 0.8522 & 0.8557 (0.0055) & -0.2054 (0.0113) & 0.8357 (0.0061) & -0.6569 (0.0101) \\ \midrule[1.25pt]
\multirow{5}{*}{\rotatebox{30}{\bvae}} & 0 & 2.5 & 0.3740  & 0.0184 & - & - & - & - & - & - & -  \\
 & 1 & 100 & 0.3807 & 0.0208 & 0.0904 (0.0091) & 0.0313 (0.0049) & 0.8457 & 0.8449 (0.0059) & -0.2010 (0.0160) & 0.7647 (0.0083) & -0.5283 (0.0125) \\
 & 10 & 2.5 & 0.3808 & 0.0858 & 0.1863 (0.0128) & 0.0644 (0.0091) & \textbf{0.9016} & 0.9039 (0.0038) & -0.2664 (0.0134) & 0.8567 (0.0053) & -0.6208 (0.0109) \\
 & 50 & 2.5 & 0.3703  & 0.1360 & 0.2314 (0.0173) & 0.0800 (0.0122) & 0.8886 & 0.8893 (0.0046) & -0.2951 (0.0122) & 0.8506 (0.0057) & -0.6414 (0.0104) \\
 & 100 & 10 & 0.3675 & 0.1476 & 0.2215 (0.0188) & 0.0770 (0.0132) & 0.8632  & 0.8681 (0.0053) & -0.2304 (0.0119) & 0.8337 (0.0062) & -0.6483 (0.0106) \\ \midrule
\multirow{5}{*}{\rotatebox{30}{\btcvae}} & 0 & 1 & 0.3897  & 0.0062 & - & - & - & - & - & - & -  \\
 & 1 & 1 & \textbf{0.3898}  & 0.0071 & -0.0123 (0.0187) & 0.0321 (0.0092) & 0.2564 & 0.2791 (0.0146) & -0.1947 (0.0140) & 0.3141 (0.0143) & -0.1164 (0.0159) \\
 & 10 & 1 & 0.3884  & 0.0087 & 0.1283 (0.0222) & 0.0991 (0.0354) & 0.8662 & 0.8735 (0.0035) & -0.3599 (0.0131) & 0.8542 (0.0042) & -0.2949 (0.0142) \\
 & 50 & 1 & 0.3876 & 0.0597 & 0.2885 (0.0327) & 0.2663 (0.0610) &  0.8584 & 0.8648 (0.0053) & -0.3617 (0.0125) & 0.9007 (0.0029) & -0.5658 (0.0138) \\
 & 100 & 1 & 0.3863 & \textbf{0.1967} & \textbf{0.3844} (0.0381)& 0.3629 (0.0657) & 0.8279 & 0.8379 (0.006) & -0.2221 (0.0110) & 0.8806 (0.0039) & -0.5880 (0.0143) \\ \bottomrule
\end{tabular}}

\end{table*}

\subsubsection{\textbf{RQ1} \RQone}\label{exp_setup_rq1}
To investigate this, we train \bvae and \bvae  with varying levels of supervision: $\{ 0\%, 1\%, 10\%, 50\%, 100\%\}$. For each combination of model, signal, level of supervision, we compute the five metrics outlined earlier. To summarize, $\dctrl$ evaluates the performance gain when a user sets a knob to its maximum setting, $\basecorr$ measures the gradual change via a correlation of manipulations and $\rmetric$; the other metrics contrast the controllability of a genre against two control genres. The holdout sets used to compute the metrics above are constructed for 100 users per genre (metrics remain constant beyond 100), while ensuring that there are at least $10/5$ items in the input/holdout sets described in Table \ref{tab:ctrl_metrics}. The number of steps taken in the latent space is 50 for computing the correlation metrics.

\subsubsection{\textbf{RQ2} \RQtwo}\label{sec:exp_setup_rq2} The relationship between disentanglement and controllability is explored in this section, where we measure if models with disentanglement ($\beta>=1$) perform better than models without ($\beta=0$). Note that $\beta=0$ implies no disentanglement, but for the \btcvae, only the TC constraint is set to 0, i.e which means $\beta=0, \gamma=\alpha=1$. In addition, we investigate if models that have higher disentanglement (MIG) scores perform better based on recommendation/ controllability metrics.  MIG is computed using code from \cite{locatello19a} \footnote{https://github.com/google-research/disentanglement\_lib/}, modified for use in \texttt{pytorch}. Since the exact M.I is intractable, the empirical M.I is computed by discretizing the values, following \cite{chen2018isolating, locatello19a}. The number of bins for the discretizer is set to 20, and it is computed for on 10,000 samples from the test set. 

\subsubsection{\textbf{RQ3} \RQthree}\label{sec:exp_setup_rq3} This question deals with the \textit{relevance} of both controlled and `default' recommendations. This is paramount, otherwise recommendations can be irrelevant. As such, NDCG@100 is computed for all models, and model selection is done on the basis of recommendation performance on the validation set. We remind the readers that $\dirrel$ has to be interpreted with $\dctrl$. That is, if $\dirrel$ is low and $\dctrl$ is high, recommended items may be personalized; however, if $\dirrel$ is higher, the models do exhibit controllability, but the recommended items might not be all personalized. Note that these metrics might be limited since a user might like these items regardless, or, such recommendations might be explicitly sought by the user. Performance for models with no disentanglement ($\beta=0$) and no supervision ($\%=0$) are also reported. 

\subsection{Experimental Setup}\label{sec:exp_setup}

This section details specifics of models being used, along with hyperparameters:

\subsubsection{Variational Distribution} Both \bvae and \btcvae use a isotropic Gaussian distribution for a prior ($p(\bz) = \Normal{\mathbf{0}, \mathbf{1}}$), and a Gaussian distribution for the variational posterior $q_\phi$:

\begin{align}
    q_\phi(\bz | \bx) = \Normal{\bz | \bmu_\phi(\bx), \operatorname{diag}(\bsigma^2_\phi(\bx))}
\end{align}

where $\bmu_\phi(\bx)$ and $\bsigma^2_\phi(\bx)$ are outputs of the encoder network. 

\subsubsection{Multinomial loss} \label{sec:mult-loss} We use the Multinomial log-likelihood for $p_\theta$, which has been shown to perform well over other losses like the Gaussian/Logistic losses \cite{liang2018variational}:

\begin{align}
    \log p_\theta(x|z) = \sum_i \bx_i \log \pi_i(\bz)
\end{align}

The final loss used for both \bvae and \btcvae methods are given in Equation \ref{eq:ss_loss}, where $\mathcal{L}_{unsup}$ is either Equation \ref{eq:bvae_loss} for the \bvae~ or Equation \ref{eq:btcvae-loss} for the \btcvae, and second term is the binary cross entropy loss (Equation \ref{eq:bce_loss}). In addition, $\gamma_{ss}=0$ for models with no supervision (\%=0), and $\gamma_{ss}=1$ otherwise (\%>0).  To obtain recommendations, items are sorted in the descending order of the likelihood predicted by the decoder.

\subsubsection{Model Hyperparameters}\label{sec:exp_hyperparams}

All models use the same neural network with a 3 layer encoder, with dimensions $|\mathcal{I}| \rightarrow 600 \rightarrow 600 \rightarrow 200$ and a single layer decoder which was found optimal for a variety of settings \cite{shenbin2020recvae}. The outputs of the encoder are two 200 dimensional real vectors representing the mean/log-variance vectors.All models were trained with Adam \cite{kingma2014adam} with a learning rate of $0.001$ and a batch size of $500$ for $200$ epochs for ML-20M and $100$ epochs for MSD. For both, we tried $\beta \in \{1, 2.5, 5, 10, 100\}$, with the `best' model selected using NDCG@100 computed on the validation set. For the \btcvae, we set $\alpha = \gamma = 1$.

\begin{table*}[htpb]
\caption{Results for MSD (Genre): Models for this dataset in particular struggle to distinguish between genres (very high CorrCtrl scores)}
\label{tab:res_msd_genre}
\resizebox{\textwidth}{!}{%
\centering
\begin{tabular}{ccc|cc|cc|c|cc|cc}
\toprule Model & \% & $\beta$ & $\uparrow$ NDCG &  MIG  & $\uparrow$ $\dctrl$ & $\downarrow$ $\dirrel$ & $\uparrow$ $\basecorr$ & $\uparrow$ $\easycorrbl$ & $\downarrow$ $\easycorrct$ & $\uparrow$ $\diffcorrbl$ & $\downarrow$ $\diffcorrct$ \\ \midrule
\multirow{3}{*}{\rotatebox{30}{\bvae}} & 0 & 0 & 0.2537  & \textbf{0.0027} & - & - & - & - & - & - & - \\
 & 50 & 0 & 0.2503  & 0.0022  & 0.0974 (0.0068) & 0.0680 (0.0119) & 0.6106 & 0.7150 (0.0119) & 0.2413 (0.0212) & 0.7251 (0.0117) & 0.2745 (0.0235) \\
 & 100 & 0 & 0.2459 & 0.0022 & 0.1378 (0.011) & 0.0972 (0.0171) & 0.6191 & 0.7173 (0.0115) & 0.2448 (0.0205) & 0.7462 (0.0110) & 0.2836 (0.0233) \\ \midrule
\multirow{3}{*}{\rotatebox{30}{\btcvae}} & 0 & 0 & \textbf{0.2620}  & 0.0022 & - & - & - & - & - & - & - \\
 & 50 & 0 & 0.2543  & 0.0022 & 0.0979 (0.0095) & 0.0715 (0.0120) & 0.6021 & 0.7122 (0.0120) & 0.2309 (0.0213) & 0.7334 (0.0116) & 0.2688 (0.0240) \\
 & 100 & 0 & 0.2509 & 0.0022 & \textbf{0.1614} (0.0118) & 0.1087 (0.0173) & \textbf{0.6522} & 0.7492 (0.0108) & 0.2612 (0.0211) & 0.7715 (0.0103) & 0.2674 (0.0243) \\ \midrule[1.5pt]
\multirow{5}{*}{\rotatebox{30}{\bvae}} & 0 & 5 & 0.2768  & 0.0055 & - & - & - & - & - & - & - \\
 & 1 & 100 & 0.2759  & 0.0076  & 0.0827 (0.0242) & 0.1535 (0.0331) & 0.5111 & 0.5659 (0.0156) & 0.2417 (0.0207) & 0.5739 (0.0159) & 0.0855 (0.0232) \\
 & 10 & 1 & 0.2745  & 0.0194 & 0.3561 (0.0527) & 0.4560 (0.0177) & 0.7034 & 0.6824 (0.0118) & 0.2903 (0.0201) & 0.6907 (0.0117) & 0.1200 (0.0231) \\
 & 50 & 10 & 0.2706  & 0.0388 & 0.4693 (0.0382) & 0.6822 (0.0202) & \textbf{0.7807} & 0.8014 (0.0048) & 0.2587 (0.0218) & 0.8003 (0.0050) & 0.1307 (0.0255) \\
 & 100 & 1 & 0.2719  & \textbf{0.4178} & \textbf{0.4853} (0.0342) & 0.7795 (0.0322) & 0.7209 & 0.7596 (0.0062) & 0.1894 (0.0222) & 0.7598 (0.0062) & 0.0613 (0.0266) \\ \midrule
\multirow{5}{*}{\rotatebox{30}{\btcvae}} & 0 & 1 & 0.2763  & 0.0145 & - & - & - & - & - & - & -  \\
 & 1 & 1 & \textbf{0.2777}  & 0.0069  & 0.0131 (0.0061) & 0.0216 (0.0085) & 0.2302 & 0.3328 (0.0195) & 0.1143 (0.0200) & 0.3352 (0.0202) & 0.1461 (0.0220) \\
 & 10 & 1 & 0.2728  & 0.0148 & 0.3461 (0.044) & 0.3915 (0.0296) & 0.7343 & 0.7501 (0.0082) & 0.2040 (0.0226) & 0.7573 (0.0082) & 0.1680 (0.0241) \\
 & 50 & 1 & 0.2675  & 0.0471 & 0.4439 (0.0366) & 0.5324 (0.0419) & 0.7820 & 0.8001 (0.0059) & 0.2405 (0.0224) & 0.8099 (0.0057) & 0.1438 (0.0259) \\
 & 100 & 1 & 0.2671 & 0.3521 & 0.4501 (0.0328) & 0.7526 (0.0405) & 0.6772 & 0.7219 (0.0073) & 0.1927 (0.021) & 0.7057 (0.0080) & 0.0539 (0.0257) \\ \bottomrule
\end{tabular}}
\end{table*}

\begin{table*}[htpb]
\centering
\caption{Results for MSD (Tag): Models confuse between tags to a lesser extent compared to Genre}
\label{tab:res_msd_tag}
\resizebox{\textwidth}{!}{%
\begin{tabular}{ccc|cc|cc|c|cc|cc}
\toprule Model & \% & $\beta$ & $\uparrow$ NDCG & MIG & $\uparrow$ $\dctrl$ & $\downarrow$ $\dirrel$ & $\uparrow$ $\basecorr$ & $\uparrow$ $\easycorrbl$ & $\downarrow$ $\easycorrct$ & $\uparrow$ $\diffcorrbl$ & $\downarrow$ $\diffcorrct$ \\ \midrule
\multirow{3}{*}{\rotatebox{30}{\bvae}} & 0 & 0 & 0.2537  & 0.0006 & - & - & - & - & - & - & - \\
 & 50 & 0 & 0.2419  & 0.0006 & 0.0832 (0.0035) & 0.0835 (0.0092) & \textbf{0.7608} & 0.7622 (0.0056) & -0.0824 (0.0068) & 0.7168 (0.0061) & -0.2136 (0.0118) \\
 & 100 & 0 & 0.2370  & 0.0006 & 0.0776 (0.0050) & 0.0790 (0.0111) & 0.6812 & 0.6804 (0.0064) & -0.0405 (0.0052) & 0.6685 (0.0065) & -0.2182 (0.0118) \\ \midrule
\multirow{3}{*}{\rotatebox{30}{\btcvae}} & 0 & 0 & \textbf{0.2620}  & 0.0006 & - & - & - & - & - & - & -  \\
 & 50 & 0 & 0.2444  & 0.0006 & \textbf{0.0905} (0.0039) & 0.0925 (0.0097) & 0.7546 & 0.7577 (0.0055) & -0.0638 (0.0058) & 0.7292 (0.0058) & -0.1664 (0.0123) \\
 & 100 & 0 & 0.2403  & 0.0006 & 0.0828 (0.0044) & 0.0807 (0.0100) & 0.6960 & 0.6989 (0.0062) & -0.0539 (0.0053) & 0.6858 (0.0062) & -0.1941 (0.0121) \\ \midrule[1.5pt]
\multirow{5}{*}{\rotatebox{30}{\bvae}} & 0 & 2.5 & 0.2768 & 0.0038 & - & - & - & - & - & - & - \\
 & 1 & 1 & 0.2749  & 0.0107 & 0.0407 (0.0095) & 0.2519 (0.0358) & 0.5815 & 0.5798 (0.0083) & -0.0865 (0.0083) & 0.5827 (0.0082) & -0.0230 (0.0115) \\
 & 10 & 1 & 0.2737  & 0.0233 & 0.1572 (0.0120) & 0.5032 (0.0310) & 0.7946 & 0.7954 (0.0041) & -0.1542 (0.0083) & 0.8074 (0.0039) & -0.1773 (0.0123) \\
 & 50 & 1 & 0.2682  & 0.0847 & 0.1886 (0.0143) & 0.5553 (0.0455) & 0.8194 & 0.8161 (0.0039) & -0.1194 (0.0075) & 0.8141 (0.0038) & -0.1912 (0.0126) \\
 & 100 & 5 & 0.2655 & 0.2244  & 0.1874 (0.0121) & 0.5763 (0.0473) & 0.8170 & 0.8192 (0.0037) & -0.1010 (0.007) & 0.8210 (0.0038) & -0.1283 (0.0130) \\ \midrule
\multirow{5}{*}{\rotatebox{30}{\btcvae}} & 0 & 1 & 0.2781  & 0.0072 & - & - & - & - & - & - & - \\
 & 1 & 1 & \textbf{0.2822}  & 0.0085 & 0.0165 (0.0042) & 0.0741 (0.0160) & 0.4415 & 0.4501 (0.0095) & -0.0359 (0.0088) & 0.4498 (0.0095) & -0.0042 (0.0112) \\
 & 10 & 1 & 0.2761  & 0.0065  & 0.0925 (0.0086) & 0.2490 (0.0404) & 0.7915 & 0.7948 (0.0048) & -0.1374 (0.0086) & 0.7872 (0.0049) & -0.0929 (0.0118) \\
 & 50 & 1 & 0.2670  & 0.1123 & 0.2018 (0.0151) & 0.4289 (0.0437) & \textbf{0.8481} & 0.8494 (0.0036) & -0.1290 (0.0084) & 0.8431 (0.0036) & -0.1133 (0.0130) \\
 & 100 & 1 & 0.2628 & \textbf{0.2272} & \textbf{0.2029} (0.0133) & 0.5042 (0.0454) & 0.8224 & 0.8224 (0.0041) & -0.0709 (0.0062) & 0.8237 (0.0039) & -0.1316 (0.0130)\\ \bottomrule
\end{tabular}%
}
\end{table*}

\section{Results}

The results of all experiments on ML-20M is reported in Table \ref{tab:res_ml20m}, on MSD (Tag) in Table \ref{tab:res_msd_genre} and finally on MSD (Genre) in Table \ref{tab:res_msd_tag}. The results indicate that while increasing supervision generally tends to increase controllability, too much supervision can sometimes harm controllability, or alternatively increase controllability at the cost of more non-personalized recommendations (Section \ref{sec:res_ss_ctrl}). The degree of supervision required for controllability varies across models/datasets/signals. In addition, adding a disentanglement objective helps across all metrics, but there appears to be no conclusive trend between degree of disentanglement and controllability (Section \ref{sec:res_disent}). Finally, controlled recommendations seem to be more personalized for ML-20M compared to MSD, where more non-personalized items get recommended (Section \ref{sec:res_recperf}). We discuss each result in detail in the sections that follow.

\subsection{Semi-Supervision and Controllability}\label{sec:res_ss_ctrl}

Surprisingly, apparent control over recommendations seems to be achieved with as \textit{little as 10\% of data} for both datasets, as measured by high $\basecorr$ and $\dctrl$. In addition, these values increase as supervision is increased to 50\% for all data and models. Supervision beyond that, however, can produce mixed results; for instance, \bvae trained on 100\% of ML-20M scores worse on $\dctrl$ and $\basecorr$. In contrast, for MSD(Genre) and MSD(Tag) (with the exception of \bvae for 100\%), $\dctrl$ increases with supervision. These trends also hold for the $\beta=0$ models. Therefore, \textit{in most cases, supervision increases apparent controllability}.

Given a model which scores high on $\dctrl$ (or $\basecorr$), finer grained $\ctrlcorr$ metrics can be analyzed, to check if changing a knob inadvertently changes other genres \footnote{Note that $\easycorrbl$ (or $\easycorrct$) cannot be compared with $\diffcorrbl$ (or $\diffcorrct$), as the inputs for computing these metrics can be very different. In addition, we observed that a lot of values of $\easycorrct$ (at user/genre level) are equal to 0.}. Increasing supervision for ML-20M makes models score better on $\diffcorrct$ (i.e more negative), while only slightly decreasing $\diffcorrbl$, indicating that supervision can help with similar genres in particular. While supervision initially improves $\easycorrct$, 100\% supervision actually worsens it slightly from 50\%. Therefore, for ML-20M, supervision can help the model avoid confusing similar genres, at the (slight) cost of controlling dissimilar genres, especially if fully supervised. 

For MSD (Genre), we first note that values of $\easycorrct$ and $\diffcorrct$ are positive for this dataset, indicating that other genres are also being manipulated. Increasing supervision increases $\easycorrbl$ and $\diffcorrbl$ i.e help control, while $\easycorrct$ and $\diffcorrct$ tend to become worse, indicating confusion among other genres. While $\easycorrbl$ and $\diffcorrbl$ becomes worse with 100\% supervision, $\easycorrct$ and $\diffcorrct$ improve, indicating that full supervision can help with confusion. In conclusion, for MSD (Genre), increasing supervision generally improves control, at the cost of other genres being also recommended, which improves with 100\% supervision for MSD(Genre). 

In MSD (Tag), supervision generally increases $\easycorrbl$ and $\diffcorrbl$. For the \btcvae, increasing supervision to 100\% makes the model score better on $\diffcorrct$, but worse on $\easycorrct$.  In contrast, for increasing supervision to 100\% for the \bvae makes the model score worse on both $\diffcorrct$ and $\easycorrct$. For the MSD (Tag) dataset, therefore, while supervision increases controllability, too much can cause confusion between tags. 

In summary, \textit{while an increase in supervision generally increases controllability} across datasets and models, and additionally \textit{reduces confusion between factors}, \textit{too much supervision can harm performance} by causing other genres to also be recommended alongside the one being controlled.

\subsection{Disentanglement and Controllability}\label{sec:res_disent}

This section discusses if disentanglement ($\beta>=1$) models are necessary, and if increased disentanglement as measured by the MIG score contributes to controllability.  

\subsubsection{Comparison with baselines} Models without disentanglement $\beta=0$, generally score worse in most respects, compared to models with disentanglement. For the ML-20M dataset, $\beta=0$ models have comparable controllability scores compared with models with disentanglement. However, this comes at a drastic reduction in recommender performance i.e the best NDCG ($0.2904$) among the baselines is worse than the lowest NDCG among models with disentanglement ($0.3675$). This is also seen for the MSD (Genre) datasets, although the performance drop is smaller. For both the MSD datasets, however, the gap observed in the controllability performance of entangled/disentangled models is much greater than the gap in the ML-20M dataset, indicating that disentanglement is necessary for controllability in these datasets. Overall, \emph{models with disentanglement perform better than models without disentanglement}. The next section compares the \emph{degree} of disentanglement with controllability.  

\subsubsection{Does increased disentanglement help?} For the ML-20M dataset, increased disentanglement results in lower recommendation (NDCG) performance. ML-20M models with high MIG scores perform better on $\dctrl$, although this doesn't seem to be necessary for \btcvae with $\%=50$, which scores high on many metrics despite having a low MIG score. For the MSD (Genre) models, a jump in MIG is accompanied by large increases in $\dctrl$ and $\dirrel$. However, as the values of MIG scores here are either high or near zero, it's difficult to make concrete conclusions about the precise relationship between disentanglement scores and controllability. We note, however, that \btcvae, which achieves better disentanglement \cite{chen2018isolating} in general, seems to perform better on the ML-20M and MSD (Tag) dataset while achieving the highest MIG scores, while \bvae performs better and achieves the highest MIG score for MSG (Genre). 

\subsection{Controlled Recommendation Performance}\label{sec:res_recperf}

We first note that \emph{as the amount of supervision, and consequently, controllability increases, average recommender performance (NDCG) decreases}. However, we argue that the benefits might outweigh the cost, as this drop is relatively small while providing control for users (or practitioners). While NDCG measures performance of non-controlled recommendations, the degree of personalization of controlled recommendations can be measured by comparing $\dctrl$ and $\dirrel$.

The results vary depending on the model or dataset. For instance, since $\dctrl$ is much higher than $\dirrel$ for \bvae with $100\%$ supervision on the ML-20M, we can conclude that it recommends personalized items instead of random items that a user might like. However, the opposite is true for \btcvae, as $\dctrl$ and $\dirrel$ are at a similar level, which means that the items recommended on controlling a genre might not be as personalized to the user. \btcvae produces the highest $\dctrl$ scores for the ML-20M and MSD(Tag) datasets, while \bvae produces the highest $\dctrl$ scores for MSD (Genre). In addition, while $\dctrl$ is higher (more items of that genre are being pushed to the top), $\dirrel$ is also very high in MSD (Genre), which mean recommendations are controllable, but might not be personalized. However, as data sparsity increases, $\dirrel$ might be overestimated because of missing relevance judgements, as  evidenced by high $\dirrel$ for almost all models trained on MSD. In addition to incompleteness, this evaluation also does not account for the exploratory/interactive/serendipitous nature of controllable recommendations. 

In summary, \textit{controlled recommendations appear to be personalized}, as evidenced by high scores of $\dctrl$, showing that users can exert control and expect personalized recommendations to an extent. However, \textit{other items of the given genre might also be recommended}, which is especially true for the MSD dataset. 

\section{Conclusion}

We showed that controllable recommendations can be achieved by leveraging the generative nature of recommenders. (Semi-)supervised disentanglement is used both to tie a generative factor to a known dimension, and to enforce disentanglement. Consequently, manipulating a dimension produces \textit{controlled} recommendations, allowing a user to express short-term/dynamic preferences, or to explore these recommendations in an interactive manner. We experimented with genre/tag distributions as supervision signals on two datasets. We proposed metrics to measure the extent of controllability in addition to the degree of personalization of controlled recommendations. Using these metrics, we showed that a user can control recommendations, while retaining personalization to an extent. We showed that such control comes only at a slight reduction in recommendation performance, but might require different degrees of supervision depending on the data, model or signals used. 

While the experiments here detail controlling only a single dimension, multiple dimensions can be manipulated allowing for greater or more nuanced control, which we leave for future work. In addition, the dimensionality of the supervision signal is limited by the dimensionality of the latent space, which limits the applicability to large sets of knobs. Evaluating the personalization of controlled recommendations presents a challenge as outlined in the previous section, which we plan to pursue in the future. In addition, this method can be used to `boost' recommendations of items belonging to a particular category, for instance, if items of category \textit{X} are being under-recommended (possibly due biases), they can be boosted to compensate for it.

\begin{acks}
The authors would like to thank Mohammad Aliannejadi, Wilker Aziz, Maurits Bleeker, Jin Huang, Antonis Krasakis, Anna Sepliarskaia, Georgios Sidiropoulos and Svitlana Vakulenko for helpful comments and feedback. The authors also thank the feedback received by reviewers during prior review processes. This research was supported by the NWO Innovational Research Incentives Scheme Vidi (016.Vidi.189.039), the NWO Smart Culture - Big Data / Digital Humanities (314-99-301), and the H2020-EU.3.4. - SOCIETAL CHALLENGES - Smart, Green And Integrated Transport (814961). All content represents the opinion of the authors, which is not necessarily shared or endorsed by their respective employers and/or sponsors.
\end{acks}

\newpage

\bibliographystyle{ACM-Reference-Format}
\bibliography{references.bib}


\begin{thebibliography}{38}


\ifx \showCODEN    \undefined \def \showCODEN     #1{\unskip}     \fi
\ifx \showDOI      \undefined \def \showDOI       #1{#1}\fi
\ifx \showISBNx    \undefined \def \showISBNx     #1{\unskip}     \fi
\ifx \showISBNxiii \undefined \def \showISBNxiii  #1{\unskip}     \fi
\ifx \showISSN     \undefined \def \showISSN      #1{\unskip}     \fi
\ifx \showLCCN     \undefined \def \showLCCN      #1{\unskip}     \fi
\ifx \shownote     \undefined \def \shownote      #1{#1}          \fi
\ifx \showarticletitle \undefined \def \showarticletitle #1{#1}   \fi
\ifx \showURL      \undefined \def \showURL       {\relax}        \fi
\providecommand\bibfield[2]{#2}
\providecommand\bibinfo[2]{#2}
\providecommand\natexlab[1]{#1}
\providecommand\showeprint[2][]{arXiv:#2}

\bibitem[\protect\citeauthoryear{Antognini and Faltings}{Antognini and
  Faltings}{2021}]%
        {antognini2021fastmulti}
\bibfield{author}{\bibinfo{person}{Diego Antognini} {and} \bibinfo{person}{Boi
  Faltings}.} \bibinfo{year}{2021}\natexlab{}.
\newblock \showarticletitle{Fast Multi-Step Critiquing for VAE-Based
  Recommender Systems}. In \bibinfo{booktitle}{\emph{Fifteenth ACM Conference
  on Recommender Systems}} (Amsterdam, Netherlands)
  \emph{(\bibinfo{series}{RecSys '21})}. \bibinfo{publisher}{Association for
  Computing Machinery}, \bibinfo{address}{New York, NY, USA},
  \bibinfo{pages}{209–219}.
\newblock
\showISBNx{9781450384582}
\urldef\tempurl%
\url{https://doi.org/10.1145/3460231.3474249}
\showDOI{\tempurl}


\bibitem[\protect\citeauthoryear{Antognini, Musat, and Faltings}{Antognini
  et~al\mbox{.}}{2020}]%
        {antognini2020trecs}
\bibfield{author}{\bibinfo{person}{Diego Antognini}, \bibinfo{person}{Claudiu
  Musat}, {and} \bibinfo{person}{Boi Faltings}.}
  \bibinfo{year}{2020}\natexlab{}.
\newblock \showarticletitle{{T-RECS:} a Transformer-based Recommender
  Generating Textual Explanations and Integrating Unsupervised Language-based
  Critiquing}.
\newblock \bibinfo{journal}{\emph{CoRR}}  \bibinfo{volume}{abs/2005.11067}
  (\bibinfo{year}{2020}).
\newblock
\showeprint[arXiv]{2005.11067}
\urldef\tempurl%
\url{https://arxiv.org/abs/2005.11067}
\showURL{%
\tempurl}


\bibitem[\protect\citeauthoryear{Bengio, Courville, and Vincent}{Bengio
  et~al\mbox{.}}{2013}]%
        {bengio2013representation}
\bibfield{author}{\bibinfo{person}{Yoshua Bengio}, \bibinfo{person}{Aaron
  Courville}, {and} \bibinfo{person}{Pascal Vincent}.}
  \bibinfo{year}{2013}\natexlab{}.
\newblock \showarticletitle{Representation learning: A review and new
  perspectives}.
\newblock \bibinfo{journal}{\emph{IEEE transactions on pattern analysis and
  machine intelligence}} \bibinfo{volume}{35}, \bibinfo{number}{8}
  (\bibinfo{year}{2013}), \bibinfo{pages}{1798--1828}.
\newblock


\bibitem[\protect\citeauthoryear{Bertin-Mahieux, Ellis, Whitman, and
  Lamere}{Bertin-Mahieux et~al\mbox{.}}{2011}]%
        {msd-Bertin-Mahieux2011}
\bibfield{author}{\bibinfo{person}{Thierry Bertin-Mahieux},
  \bibinfo{person}{Daniel~P.W. Ellis}, \bibinfo{person}{Brian Whitman}, {and}
  \bibinfo{person}{Paul Lamere}.} \bibinfo{year}{2011}\natexlab{}.
\newblock \showarticletitle{The Million Song Dataset}. In
  \bibinfo{booktitle}{\emph{{Proceedings of the 12th International Conference
  on Music Information Retrieval ({ISMIR} 2011)}}}.
\newblock


\bibitem[\protect\citeauthoryear{Blei, Kucukelbir, and McAuliffe}{Blei
  et~al\mbox{.}}{2017}]%
        {blei2017variational}
\bibfield{author}{\bibinfo{person}{David~M Blei}, \bibinfo{person}{Alp
  Kucukelbir}, {and} \bibinfo{person}{Jon~D McAuliffe}.}
  \bibinfo{year}{2017}\natexlab{}.
\newblock \showarticletitle{Variational inference: A review for statisticians}.
\newblock \bibinfo{journal}{\emph{Journal of the American statistical
  Association}} \bibinfo{volume}{112}, \bibinfo{number}{518}
  (\bibinfo{year}{2017}), \bibinfo{pages}{859--877}.
\newblock


\bibitem[\protect\citeauthoryear{Burgess, Higgins, Pal, Matthey, Watters,
  Desjardins, and Lerchner}{Burgess et~al\mbox{.}}{2018}]%
        {burgess2018understanding}
\bibfield{author}{\bibinfo{person}{Christopher~P Burgess},
  \bibinfo{person}{Irina Higgins}, \bibinfo{person}{Arka Pal},
  \bibinfo{person}{Loic Matthey}, \bibinfo{person}{Nick Watters},
  \bibinfo{person}{Guillaume Desjardins}, {and} \bibinfo{person}{Alexander
  Lerchner}.} \bibinfo{year}{2018}\natexlab{}.
\newblock \showarticletitle{Understanding disentangling in $beta$-VAE}.
\newblock \bibinfo{journal}{\emph{arXiv preprint arXiv:1804.03599}}
  (\bibinfo{year}{2018}).
\newblock


\bibitem[\protect\citeauthoryear{Cen, Zhang, Zou, Zhou, Yang, and Tang}{Cen
  et~al\mbox{.}}{2020}]%
        {cen2020controllable}
\bibfield{author}{\bibinfo{person}{Yukuo Cen}, \bibinfo{person}{Jianwei Zhang},
  \bibinfo{person}{Xu Zou}, \bibinfo{person}{Chang Zhou},
  \bibinfo{person}{Hongxia Yang}, {and} \bibinfo{person}{Jie Tang}.}
  \bibinfo{year}{2020}\natexlab{}.
\newblock \showarticletitle{Controllable Multi-Interest Framework for
  Recommendation}. In \bibinfo{booktitle}{\emph{Proceedings of the 26th ACM
  SIGKDD International Conference on Knowledge Discovery \& Data Mining}}.
  \bibinfo{pages}{2942--2951}.
\newblock


\bibitem[\protect\citeauthoryear{Chen, Li, Grosse, and Duvenaud}{Chen
  et~al\mbox{.}}{2018}]%
        {chen2018isolating}
\bibfield{author}{\bibinfo{person}{Ricky~TQ Chen}, \bibinfo{person}{Xuechen
  Li}, \bibinfo{person}{Roger~B Grosse}, {and} \bibinfo{person}{David~K
  Duvenaud}.} \bibinfo{year}{2018}\natexlab{}.
\newblock \showarticletitle{Isolating sources of disentanglement in variational
  autoencoders}. In \bibinfo{booktitle}{\emph{Advances in Neural Information
  Processing Systems}}. \bibinfo{pages}{2610--2620}.
\newblock


\bibitem[\protect\citeauthoryear{Cui, Yu, Wu, Liu, and Wang}{Cui
  et~al\mbox{.}}{2020}]%
        {cui2020disentangled}
\bibfield{author}{\bibinfo{person}{Zeyu Cui}, \bibinfo{person}{Feng Yu},
  \bibinfo{person}{Shu Wu}, \bibinfo{person}{Qiang Liu}, {and}
  \bibinfo{person}{Liang Wang}.} \bibinfo{year}{2020}\natexlab{}.
\newblock \showarticletitle{Disentangled Item Representation for Recommender
  Systems}.
\newblock \bibinfo{journal}{\emph{arXiv preprint arXiv:2008.07178}}
  (\bibinfo{year}{2020}).
\newblock


\bibitem[\protect\citeauthoryear{Harper and Konstan}{Harper and
  Konstan}{2015}]%
        {movielens}
\bibfield{author}{\bibinfo{person}{F.~Maxwell Harper} {and}
  \bibinfo{person}{Joseph~A. Konstan}.} \bibinfo{year}{2015}\natexlab{}.
\newblock \showarticletitle{The MovieLens Datasets: History and Context}.
\newblock \bibinfo{journal}{\emph{ACM Trans. Interact. Intell. Syst.}}
  \bibinfo{volume}{5}, \bibinfo{number}{4}, Article \bibinfo{articleno}{19}
  (\bibinfo{date}{Dec.} \bibinfo{year}{2015}), \bibinfo{numpages}{19}~pages.
\newblock
\showISSN{2160-6455}
\urldef\tempurl%
\url{https://doi.org/10.1145/2827872}
\showDOI{\tempurl}


\bibitem[\protect\citeauthoryear{He, Liao, Zhang, Nie, Hu, and Chua}{He
  et~al\mbox{.}}{2017}]%
        {he2017neural}
\bibfield{author}{\bibinfo{person}{Xiangnan He}, \bibinfo{person}{Lizi Liao},
  \bibinfo{person}{Hanwang Zhang}, \bibinfo{person}{Liqiang Nie},
  \bibinfo{person}{Xia Hu}, {and} \bibinfo{person}{Tat-Seng Chua}.}
  \bibinfo{year}{2017}\natexlab{}.
\newblock \showarticletitle{Neural collaborative filtering}. In
  \bibinfo{booktitle}{\emph{Proceedings of the 26th international conference on
  world wide web}}. \bibinfo{pages}{173--182}.
\newblock


\bibitem[\protect\citeauthoryear{Higgins, Matthey, Pal, Burgess, Glorot,
  Botvinick, Mohamed, and Lerchner}{Higgins et~al\mbox{.}}{2017}]%
        {higgins2017beta}
\bibfield{author}{\bibinfo{person}{Irina Higgins}, \bibinfo{person}{Loic
  Matthey}, \bibinfo{person}{Arka Pal}, \bibinfo{person}{Christopher Burgess},
  \bibinfo{person}{Xavier Glorot}, \bibinfo{person}{Matthew Botvinick},
  \bibinfo{person}{Shakir Mohamed}, {and} \bibinfo{person}{Alexander
  Lerchner}.} \bibinfo{year}{2017}\natexlab{}.
\newblock \showarticletitle{beta-VAE: Learning Basic Visual Concepts with a
  Constrained Variational Framework.}
\newblock \bibinfo{journal}{\emph{Iclr}} \bibinfo{volume}{2},
  \bibinfo{number}{5} (\bibinfo{year}{2017}), \bibinfo{pages}{6}.
\newblock


\bibitem[\protect\citeauthoryear{Jannach, Manzoor, Cai, and Chen}{Jannach
  et~al\mbox{.}}{2020}]%
        {jannach2020survey}
\bibfield{author}{\bibinfo{person}{Dietmar Jannach}, \bibinfo{person}{Ahtsham
  Manzoor}, \bibinfo{person}{Wanling Cai}, {and} \bibinfo{person}{Li Chen}.}
  \bibinfo{year}{2020}\natexlab{}.
\newblock \showarticletitle{A survey on conversational recommender systems}.
\newblock \bibinfo{journal}{\emph{arXiv preprint arXiv:2004.00646}}
  (\bibinfo{year}{2020}).
\newblock


\bibitem[\protect\citeauthoryear{Jordan, Ghahramani, Jaakkola, and Saul}{Jordan
  et~al\mbox{.}}{1999}]%
        {jordan1999introduction}
\bibfield{author}{\bibinfo{person}{Michael~I Jordan}, \bibinfo{person}{Zoubin
  Ghahramani}, \bibinfo{person}{Tommi~S Jaakkola}, {and}
  \bibinfo{person}{Lawrence~K Saul}.} \bibinfo{year}{1999}\natexlab{}.
\newblock \showarticletitle{An introduction to variational methods for
  graphical models}.
\newblock \bibinfo{journal}{\emph{Machine learning}} \bibinfo{volume}{37},
  \bibinfo{number}{2} (\bibinfo{year}{1999}), \bibinfo{pages}{183--233}.
\newblock


\bibitem[\protect\citeauthoryear{Kim and Mnih}{Kim and Mnih}{2018}]%
        {kim2018disentangling}
\bibfield{author}{\bibinfo{person}{Hyunjik Kim} {and} \bibinfo{person}{Andriy
  Mnih}.} \bibinfo{year}{2018}\natexlab{}.
\newblock \showarticletitle{Disentangling by factorising}.
\newblock \bibinfo{journal}{\emph{arXiv preprint arXiv:1802.05983}}
  (\bibinfo{year}{2018}).
\newblock


\bibitem[\protect\citeauthoryear{Kingma and Ba}{Kingma and Ba}{2014}]%
        {kingma2014adam}
\bibfield{author}{\bibinfo{person}{Diederik~P Kingma} {and}
  \bibinfo{person}{Jimmy Ba}.} \bibinfo{year}{2014}\natexlab{}.
\newblock \showarticletitle{Adam: A method for stochastic optimization}.
\newblock \bibinfo{journal}{\emph{arXiv preprint arXiv:1412.6980}}
  (\bibinfo{year}{2014}).
\newblock


\bibitem[\protect\citeauthoryear{Kingma and Welling}{Kingma and
  Welling}{2013}]%
        {kingma2013auto}
\bibfield{author}{\bibinfo{person}{Diederik~P Kingma} {and}
  \bibinfo{person}{Max Welling}.} \bibinfo{year}{2013}\natexlab{}.
\newblock \showarticletitle{Auto-encoding variational bayes}.
\newblock \bibinfo{journal}{\emph{arXiv preprint arXiv:1312.6114}}
  (\bibinfo{year}{2013}).
\newblock


\bibitem[\protect\citeauthoryear{Li, Sanner, Luo, and Wu}{Li
  et~al\mbox{.}}{2020}]%
        {li2020rankopt}
\bibfield{author}{\bibinfo{person}{Hanze Li}, \bibinfo{person}{Scott Sanner},
  \bibinfo{person}{Kai Luo}, {and} \bibinfo{person}{Ga Wu}.}
  \bibinfo{year}{2020}\natexlab{}.
\newblock \showarticletitle{A Ranking Optimization Approach to Latent Linear
  Critiquing for Conversational Recommender Systems}. In
  \bibinfo{booktitle}{\emph{Fourteenth ACM Conference on Recommender Systems}}
  (Virtual Event, Brazil) \emph{(\bibinfo{series}{RecSys '20})}.
  \bibinfo{publisher}{Association for Computing Machinery},
  \bibinfo{address}{New York, NY, USA}, \bibinfo{pages}{13–22}.
\newblock
\showISBNx{9781450375832}
\urldef\tempurl%
\url{https://doi.org/10.1145/3383313.3412240}
\showDOI{\tempurl}


\bibitem[\protect\citeauthoryear{Liang, Krishnan, Hoffman, and Jebara}{Liang
  et~al\mbox{.}}{2018}]%
        {liang2018variational}
\bibfield{author}{\bibinfo{person}{Dawen Liang}, \bibinfo{person}{Rahul~G
  Krishnan}, \bibinfo{person}{Matthew~D Hoffman}, {and} \bibinfo{person}{Tony
  Jebara}.} \bibinfo{year}{2018}\natexlab{}.
\newblock \showarticletitle{Variational autoencoders for collaborative
  filtering}. In \bibinfo{booktitle}{\emph{Proceedings of the 2018 World Wide
  Web Conference}}. \bibinfo{pages}{689--698}.
\newblock


\bibitem[\protect\citeauthoryear{Locatello, Bauer, Lucic, Raetsch, Gelly,
  Sch{\"o}lkopf, and Bachem}{Locatello et~al\mbox{.}}{2019a}]%
        {locatello19a}
\bibfield{author}{\bibinfo{person}{Francesco Locatello},
  \bibinfo{person}{Stefan Bauer}, \bibinfo{person}{Mario Lucic},
  \bibinfo{person}{Gunnar Raetsch}, \bibinfo{person}{Sylvain Gelly},
  \bibinfo{person}{Bernhard Sch{\"o}lkopf}, {and} \bibinfo{person}{Olivier
  Bachem}.} \bibinfo{year}{2019}\natexlab{a}.
\newblock \showarticletitle{Challenging Common Assumptions in the Unsupervised
  Learning of Disentangled Representations} \emph{(\bibinfo{series}{Proceedings
  of Machine Learning Research}, Vol.~\bibinfo{volume}{97})},
  \bibfield{editor}{\bibinfo{person}{Kamalika Chaudhuri} {and}
  \bibinfo{person}{Ruslan Salakhutdinov}} (Eds.). \bibinfo{publisher}{PMLR},
  \bibinfo{address}{Long Beach, California, USA}, \bibinfo{pages}{4114--4124}.
\newblock
\urldef\tempurl%
\url{http://proceedings.mlr.press/v97/locatello19a.html}
\showURL{%
\tempurl}


\bibitem[\protect\citeauthoryear{Locatello, Tschannen, Bauer, R{\"a}tsch,
  Sch{\"o}lkopf, and Bachem}{Locatello et~al\mbox{.}}{2019b}]%
        {locatello2019disentangling}
\bibfield{author}{\bibinfo{person}{Francesco Locatello},
  \bibinfo{person}{Michael Tschannen}, \bibinfo{person}{Stefan Bauer},
  \bibinfo{person}{Gunnar R{\"a}tsch}, \bibinfo{person}{Bernhard
  Sch{\"o}lkopf}, {and} \bibinfo{person}{Olivier Bachem}.}
  \bibinfo{year}{2019}\natexlab{b}.
\newblock \showarticletitle{Disentangling factors of variation using few
  labels}.
\newblock \bibinfo{journal}{\emph{arXiv preprint arXiv:1905.01258}}
  (\bibinfo{year}{2019}).
\newblock


\bibitem[\protect\citeauthoryear{Luo, Sanner, Wu, Li, and Yang}{Luo
  et~al\mbox{.}}{2020a}]%
        {luo202latentlinear}
\bibfield{author}{\bibinfo{person}{Kai Luo}, \bibinfo{person}{Scott Sanner},
  \bibinfo{person}{Ga Wu}, \bibinfo{person}{Hanze Li}, {and}
  \bibinfo{person}{Hojin Yang}.} \bibinfo{year}{2020}\natexlab{a}.
\newblock \showarticletitle{Latent Linear Critiquing for Conversational
  Recommender Systems}. In \bibinfo{booktitle}{\emph{Proceedings of The Web
  Conference 2020}} (Taipei, Taiwan) \emph{(\bibinfo{series}{WWW '20})}.
  \bibinfo{publisher}{Association for Computing Machinery},
  \bibinfo{address}{New York, NY, USA}, \bibinfo{pages}{2535–2541}.
\newblock
\showISBNx{9781450370233}
\urldef\tempurl%
\url{https://doi.org/10.1145/3366423.3380003}
\showDOI{\tempurl}


\bibitem[\protect\citeauthoryear{Luo, Yang, Wu, and Sanner}{Luo
  et~al\mbox{.}}{2020b}]%
        {luo2020deepcritiquing}
\bibfield{author}{\bibinfo{person}{Kai Luo}, \bibinfo{person}{Hojin Yang},
  \bibinfo{person}{Ga Wu}, {and} \bibinfo{person}{Scott Sanner}.}
  \bibinfo{year}{2020}\natexlab{b}.
\newblock \bibinfo{booktitle}{\emph{Deep Critiquing for VAE-Based Recommender
  Systems}}.
\newblock \bibinfo{publisher}{Association for Computing Machinery},
  \bibinfo{address}{New York, NY, USA}, \bibinfo{pages}{1269–1278}.
\newblock
\showISBNx{9781450380164}
\urldef\tempurl%
\url{https://doi.org/10.1145/3397271.3401091}
\showURL{%
\tempurl}


\bibitem[\protect\citeauthoryear{Ma, Cui, Kuang, Wang, and Zhu}{Ma
  et~al\mbox{.}}{2019a}]%
        {ma2019disentangled}
\bibfield{author}{\bibinfo{person}{Jianxin Ma}, \bibinfo{person}{Peng Cui},
  \bibinfo{person}{Kun Kuang}, \bibinfo{person}{Xin Wang}, {and}
  \bibinfo{person}{Wenwu Zhu}.} \bibinfo{year}{2019}\natexlab{a}.
\newblock \showarticletitle{Disentangled graph convolutional networks}. In
  \bibinfo{booktitle}{\emph{International Conference on Machine Learning}}.
  \bibinfo{pages}{4212--4221}.
\newblock


\bibitem[\protect\citeauthoryear{Ma, Zhou, Cui, Yang, and Zhu}{Ma
  et~al\mbox{.}}{2019b}]%
        {ma2019learning}
\bibfield{author}{\bibinfo{person}{Jianxin Ma}, \bibinfo{person}{Chang Zhou},
  \bibinfo{person}{Peng Cui}, \bibinfo{person}{Hongxia Yang}, {and}
  \bibinfo{person}{Wenwu Zhu}.} \bibinfo{year}{2019}\natexlab{b}.
\newblock \showarticletitle{Learning disentangled representations for
  recommendation}. In \bibinfo{booktitle}{\emph{Advances in Neural Information
  Processing Systems}}. \bibinfo{pages}{5712--5723}.
\newblock


\bibitem[\protect\citeauthoryear{Ma, Zhou, Yang, Cui, Wang, and Zhu}{Ma
  et~al\mbox{.}}{2020}]%
        {ma2020disentangled}
\bibfield{author}{\bibinfo{person}{Jianxin Ma}, \bibinfo{person}{Chang Zhou},
  \bibinfo{person}{Hongxia Yang}, \bibinfo{person}{Peng Cui},
  \bibinfo{person}{Xin Wang}, {and} \bibinfo{person}{Wenwu Zhu}.}
  \bibinfo{year}{2020}\natexlab{}.
\newblock \showarticletitle{Disentangled Self-Supervision in Sequential
  Recommenders}. In \bibinfo{booktitle}{\emph{Proceedings of the 26th ACM
  SIGKDD International Conference on Knowledge Discovery \& Data Mining}}.
  \bibinfo{pages}{483--491}.
\newblock


\bibitem[\protect\citeauthoryear{Matthey, Higgins, Hassabis, and
  Lerchner}{Matthey et~al\mbox{.}}{2017}]%
        {dsprites17}
\bibfield{author}{\bibinfo{person}{Loic Matthey}, \bibinfo{person}{Irina
  Higgins}, \bibinfo{person}{Demis Hassabis}, {and} \bibinfo{person}{Alexander
  Lerchner}.} \bibinfo{year}{2017}\natexlab{}.
\newblock \bibinfo{title}{dSprites: Disentanglement testing Sprites dataset}.
\newblock
  \bibinfo{howpublished}{https://github.com/deepmind/dsprites-dataset/}.
\newblock


\bibitem[\protect\citeauthoryear{Nema, Karatzoglou, and Radlinski}{Nema
  et~al\mbox{.}}{2021}]%
        {nema2021disentpref}
\bibfield{author}{\bibinfo{person}{Preksha Nema}, \bibinfo{person}{Alexandros
  Karatzoglou}, {and} \bibinfo{person}{Filip Radlinski}.}
  \bibinfo{year}{2021}\natexlab{}.
\newblock \showarticletitle{Disentangling Preference Representations for
  Recommendation Critiquing with $\beta$-VAE}. In
  \bibinfo{booktitle}{\emph{30th ACM International Conference on Information
  and Knowledge Management (CIKM 2021)}}. \bibinfo{address}{New York}.
\newblock


\bibitem[\protect\citeauthoryear{Ricci, Rokach, Shapira, and Kantor}{Ricci
  et~al\mbox{.}}{2010}]%
        {riccihandbook}
\bibfield{author}{\bibinfo{person}{Francesco Ricci}, \bibinfo{person}{Lior
  Rokach}, \bibinfo{person}{Bracha Shapira}, {and} \bibinfo{person}{Paul~B.
  Kantor}.} \bibinfo{year}{2010}\natexlab{}.
\newblock \bibinfo{booktitle}{\emph{Recommender Systems Handbook}
  (\bibinfo{edition}{1st} ed.)}.
\newblock \bibinfo{publisher}{Springer-Verlag}, \bibinfo{address}{Berlin,
  Heidelberg}.
\newblock
\showISBNx{0387858199}


\bibitem[\protect\citeauthoryear{Shenbin, Alekseev, Tutubalina, Malykh, and
  Nikolenko}{Shenbin et~al\mbox{.}}{2020}]%
        {shenbin2020recvae}
\bibfield{author}{\bibinfo{person}{Ilya Shenbin}, \bibinfo{person}{Anton
  Alekseev}, \bibinfo{person}{Elena Tutubalina}, \bibinfo{person}{Valentin
  Malykh}, {and} \bibinfo{person}{Sergey~I Nikolenko}.}
  \bibinfo{year}{2020}\natexlab{}.
\newblock \showarticletitle{RecVAE: A New Variational Autoencoder for Top-N
  Recommendations with Implicit Feedback}. In
  \bibinfo{booktitle}{\emph{Proceedings of the 13th International Conference on
  Web Search and Data Mining}}. \bibinfo{pages}{528--536}.
\newblock


\bibitem[\protect\citeauthoryear{Wang, Wang, and Yeung}{Wang
  et~al\mbox{.}}{2015}]%
        {wang2015collaborative}
\bibfield{author}{\bibinfo{person}{Hao Wang}, \bibinfo{person}{Naiyan Wang},
  {and} \bibinfo{person}{Dit-Yan Yeung}.} \bibinfo{year}{2015}\natexlab{}.
\newblock \showarticletitle{Collaborative deep learning for recommender
  systems}. In \bibinfo{booktitle}{\emph{Proceedings of the 21th ACM SIGKDD
  international conference on knowledge discovery and data mining}}.
  \bibinfo{pages}{1235--1244}.
\newblock


\bibitem[\protect\citeauthoryear{Wang, Zhou, Yang, Yang, and He}{Wang
  et~al\mbox{.}}{2021}]%
        {wang2021controllable}
\bibfield{author}{\bibinfo{person}{Haonan Wang}, \bibinfo{person}{Chang Zhou},
  \bibinfo{person}{Carl Yang}, \bibinfo{person}{Hongxia Yang}, {and}
  \bibinfo{person}{Jingrui He}.} \bibinfo{year}{2021}\natexlab{}.
\newblock \showarticletitle{Controllable Gradient Item Retrieval}. In
  \bibinfo{booktitle}{\emph{Proceedings of the Web Conference 2021}}
  (Ljubljana, Slovenia) \emph{(\bibinfo{series}{WWW '21})}.
  \bibinfo{publisher}{Association for Computing Machinery},
  \bibinfo{address}{New York, NY, USA}, \bibinfo{pages}{768–777}.
\newblock
\showISBNx{9781450383127}
\urldef\tempurl%
\url{https://doi.org/10.1145/3442381.3449963}
\showDOI{\tempurl}


\bibitem[\protect\citeauthoryear{Wang, He, Wang, Feng, and Chua}{Wang
  et~al\mbox{.}}{2019}]%
        {wang2019neural}
\bibfield{author}{\bibinfo{person}{Xiang Wang}, \bibinfo{person}{Xiangnan He},
  \bibinfo{person}{Meng Wang}, \bibinfo{person}{Fuli Feng}, {and}
  \bibinfo{person}{Tat-Seng Chua}.} \bibinfo{year}{2019}\natexlab{}.
\newblock \showarticletitle{Neural graph collaborative filtering}. In
  \bibinfo{booktitle}{\emph{Proceedings of the 42nd international ACM SIGIR
  conference on Research and development in Information Retrieval}}.
  \bibinfo{pages}{165--174}.
\newblock


\bibitem[\protect\citeauthoryear{Wang, Jin, Zhang, He, Xu, and Chua}{Wang
  et~al\mbox{.}}{2020}]%
        {wang2020disentangled}
\bibfield{author}{\bibinfo{person}{Xiang Wang}, \bibinfo{person}{Hongye Jin},
  \bibinfo{person}{An Zhang}, \bibinfo{person}{Xiangnan He},
  \bibinfo{person}{Tong Xu}, {and} \bibinfo{person}{Tat-Seng Chua}.}
  \bibinfo{year}{2020}\natexlab{}.
\newblock \showarticletitle{Disentangled Graph Collaborative Filtering}. In
  \bibinfo{booktitle}{\emph{Proceedings of the 43rd International ACM SIGIR
  Conference on Research and Development in Information Retrieval}}.
  \bibinfo{pages}{1001--1010}.
\newblock


\bibitem[\protect\citeauthoryear{Wu, Luo, Sanner, and Soh}{Wu
  et~al\mbox{.}}{2019}]%
        {gao2019dlc}
\bibfield{author}{\bibinfo{person}{Ga Wu}, \bibinfo{person}{Kai Luo},
  \bibinfo{person}{Scott Sanner}, {and} \bibinfo{person}{Harold Soh}.}
  \bibinfo{year}{2019}\natexlab{}.
\newblock \showarticletitle{Deep Language-Based Critiquing for Recommender
  Systems}. In \bibinfo{booktitle}{\emph{Proceedings of the 13th ACM Conference
  on Recommender Systems}} (Copenhagen, Denmark) \emph{(\bibinfo{series}{RecSys
  '19})}. \bibinfo{publisher}{Association for Computing Machinery},
  \bibinfo{address}{New York, NY, USA}, \bibinfo{pages}{137–145}.
\newblock
\showISBNx{9781450362436}
\urldef\tempurl%
\url{https://doi.org/10.1145/3298689.3347009}
\showDOI{\tempurl}


\bibitem[\protect\citeauthoryear{Wu, DuBois, Zheng, and Ester}{Wu
  et~al\mbox{.}}{2016}]%
        {wu2016collaborative}
\bibfield{author}{\bibinfo{person}{Yao Wu}, \bibinfo{person}{Christopher
  DuBois}, \bibinfo{person}{Alice~X Zheng}, {and} \bibinfo{person}{Martin
  Ester}.} \bibinfo{year}{2016}\natexlab{}.
\newblock \showarticletitle{Collaborative denoising auto-encoders for top-n
  recommender systems}. In \bibinfo{booktitle}{\emph{Proceedings of the Ninth
  ACM International Conference on Web Search and Data Mining}}.
  \bibinfo{pages}{153--162}.
\newblock


\bibitem[\protect\citeauthoryear{Yang, Shen, and Sanner}{Yang
  et~al\mbox{.}}{2021}]%
        {yang2021bayesiancrit}
\bibfield{author}{\bibinfo{person}{Hojin Yang}, \bibinfo{person}{Tianshu Shen},
  {and} \bibinfo{person}{Scott Sanner}.} \bibinfo{year}{2021}\natexlab{}.
\newblock \bibinfo{booktitle}{\emph{Bayesian Critiquing with Keyphrase
  Activation Vectors for VAE-Based Recommender Systems}}.
\newblock \bibinfo{publisher}{Association for Computing Machinery},
  \bibinfo{address}{New York, NY, USA}, \bibinfo{pages}{2111–2115}.
\newblock
\showISBNx{9781450380379}
\urldef\tempurl%
\url{https://doi.org/10.1145/3404835.3463108}
\showURL{%
\tempurl}


\bibitem[\protect\citeauthoryear{Zhang, Yao, Sun, and Tay}{Zhang
  et~al\mbox{.}}{2019}]%
        {zhang2019deep}
\bibfield{author}{\bibinfo{person}{Shuai Zhang}, \bibinfo{person}{Lina Yao},
  \bibinfo{person}{Aixin Sun}, {and} \bibinfo{person}{Yi Tay}.}
  \bibinfo{year}{2019}\natexlab{}.
\newblock \showarticletitle{Deep learning based recommender system: A survey
  and new perspectives}.
\newblock \bibinfo{journal}{\emph{ACM Computing Surveys (CSUR)}}
  \bibinfo{volume}{52}, \bibinfo{number}{1} (\bibinfo{year}{2019}),
  \bibinfo{pages}{1--38}.
\newblock


\end{thebibliography}

\end{document}